\documentclass[proof]{WileyASNA-v1}

\articletype{Article Type}%

\received{<day> <Month>, <year>}
\revised{<day> <Month>, <year>}
\accepted{<day> <Month>, <year>}

\newcommand{\ledd}{$\lambda_\mathrm{Edd}$}

\begin{document}

\title{X-HESS: a large sample of highly accreting serendipitous AGN under the \textit{XMM-Newton} microscope}

\author[1,2]{M. Laurenti}

\author[1,3,4,5]{F. Tombesi}

\author[1,6]{F. Vagnetti}

\author[3]{E. Piconcelli}

\author[7]{M. Guainazzi}

\author[2,3]{R. Middei}

\authormark{M. Laurenti \textsc{et al}}

\address[1]{Dipartimento di Fisica, Università di Roma “Tor Vergata”, Via della Ricerca Scientifica 1, 00133 Roma, Italy}

\address[2]{Space Science Data Center, SSDC, ASI, Via del Politecnico snc, 00133 Roma, Italy}

\address[3]{INAF – Osservatorio Astronomico di Roma, Via Frascati 33, 00040 Monte Porzio Catone, Italy}

\address[4]{Department of Astronomy, University of Maryland, College Park, MD 20742, USA}

\address[5]{NASA/Goddard Space Flight Center, Greenbelt, MD 20771, USA}

\address[6]{INAF – Istituto di Astrofisica e Planetologia Spaziali, Via del Fosso del Caveliere 100, 00133 Roma, Italy}

\address[7]{ESA European Space Research and Technology Centre (ESTEC), Keplerlaan 1, 2201 AZ, Noordwijk, The Netherlands}

\corres{\email{marco.laurenti@roma2.infn.it}}

%\presentaddress{}

\abstract[Abstract]{The bulk of X-ray spectroscopic studies of active galactic nuclei (AGN) are focused on local ($z < 0.1$) sources with low-to-moderate ($< 0.3$) Eddington ratio ($\lambda_\mathrm{Edd}$). It is then mandatory to overcome this limitation and improve our understanding of highly accreting AGN. In this work we present the preliminary results from the analysis of a sample of $\sim70$ high-$\lambda_\mathrm{Edd}$ radio-quiet AGN at $0.06 \leq z \leq 3.3$, based on the 10th release of the \textit{XMM-Newton} serendipitous source catalogue, that we named as \textit{XMM-Newton} High-Eddington Serendipitous AGN Sample (X-HESS). Almost $\sim35\%$ of the X-HESS AGN have multi-epoch archival observations and $\sim70\%$ of the sources can rely on simultaneous OM optical data. 
First results reveal sources showing signatures of ultra-fast outflows and remarkable long- and short-term X-ray flux variations. Indeed in J095847.88+690532.7 ($z \sim 1.3$), one of the most densely monitored objects hosting a $\sim$$10^9\,M_\odot$ supermassive black hole, we discovered a variation of the soft X-ray flux by a factor of > 2 over approximately one week (rest-frame). Large variations in the power-law continuum photon index $\Gamma$ are also observed, questioning expectations from previously reported $\Gamma - \lambda_\mathrm{Edd}$ relations, for which $\Gamma \geq 2$ would be a ubiquitous hallmark of AGN with $\lambda_\mathrm{Edd} \sim 1$.}


\jnlcitation{ \cname{%
\author{M. Laurenti},
\author{F. Tombesi}, 
\author{F. Vagnetti}, 
\author{E. Piconcelli},
\author{M. Guainazzi}, and
\author{R. Middei}} (\cyear{2023}), 
\ctitle{X-HESS: a large sample of highly accreting serendipitous AGN under the \textit{XMM-Newton} microscope}, \cjournal{}, \cvol{}.}

\maketitle

\section{Introduction}\label{intro}

Active galactic nuclei (AGN) are very energetic extragalactic sources powered by gas accretion onto a central supermassive black hole (SMBH). A well known proxy for the AGN accretion activity is represented by the Eddington ratio parameter, $\lambda_\mathrm{Edd} = L_\mathrm{bol}/L_\mathrm{Edd}$, where $L_\mathrm{bol}$ and $L_\mathrm{Edd}$ indicate the bolometric and Eddington luminosity, respectively.

The bulk of X-ray spectroscopic studies of AGN are focused on local ($z < 0.1$) sources with low-to-moderate ($< 0.3$) Eddington ratio \citep[e.g.][]{nandra1994, piconcelli2005, bianchi2009, ricci2018}. 
Highly accreting AGN ($\lambda_\mathrm{Edd} \gtrsim 0.5$) have often been overlooked, also as a consequence of their relative paucity in the local Universe, with the notable exception of Narrow Line Seyfert 1 galaxies \citep[NLSy1s; e.g.][]{brandt1997, gallo2006, costantini2007, jin2013, fabian2013, waddell2020}.
However NLSy1s might represent a biased population, since they consist in a peculiar and restricted class of AGN hosting only low-mass SMBH ($M_\mathrm{BH} \sim 10^6-10^7\,M_\odot$). 
Thus it would be desirable to expand the analysis of high-\ledd\ AGN by including sources distributed among a broader redshift interval and hosting more massive SMBHs.
Indeed, a deeper understanding of the physical processes occurring in the whole population of highly accreting AGN would be extremely interesting for several reasons. 
The standard model of accretion discs, developed by \citet{shakura1973}, predicts an optically thick, geometrically thin disc that is radiatively efficient, and AGN accreting at low and moderate Eddington ratios are consistent with this scheme. However, radiation pressure gains increasing importance at higher \ledd\ and eventually contributes to thicken the innermost regions of the accretion disc along the vertical scale.
This leads to a system that is both optically and geometrically thick, also known as slim disc \citep[e.g.][]{abramowicz1988, watarai2000, chen2004, sadowski2011}. 
The accretion flow in the slim disc is characterised by frequent interactions between matter and radiation, causing a significant increase of the diffusion timescale of the photons, which may easily exceed the accretion timescale. As a result, the radiation energy is trapped in the flow and advected inwards \citep[e.g.][]{mineshige2000, oshuga2002}.    
Photon trapping effects imply that the observed disc luminosity ($\sim L_\mathrm{bol}$) and, in turn, \ledd\ are expected to saturate to a limiting value of approximately 5$-$10 $L_\mathrm{Edd}$ for steadily increasing accretion rates, while the effective temperature profile of the disc flattens and is no longer proportional to $r^{-3/4}$, as in the standard discs.

Moreover, the study of high-\ledd\ AGN is important in terms of their cosmological implications. In the past few years there has been an increasing effort to detect quasars (i.e. AGN with $L_\mathrm{bol} > 10^{46}$ erg s$^{-1}$ ; QSOs hereafter) shining at $z \sim6$$-$$7$, when the Universe was less than 1 Gyr old \citep[e.g.][]{Wu2015, banados2016, mazzucchelli2017}.
Since the vast majority of these QSOs host massive SMBH ($\sim 10^9\,M_\odot$), this leads to face the difficult task to comprehend the nature of the physical processes that allowed the black hole seeds to grow up to a billion solar masses in such a relatively short amount of time.
Several scenarios have been proposed (see e.g. the review in \cite{valiante2017}), and it has been suggested that these SMBHs may evolve via gas accretion at a rate equal to or above the Eddington rate.

Finally, a high-\ledd\ is invoked as a key ingredient for the launching of powerful nuclear outflows \citep[e.g.][]{proga2005, zubovas2013, king2015}, such as warm absorbers and ultra-fast outflows (UFOs) that could be capable of controlling the growth and evolution of the host galaxy \citep[e.g.][]{hopkins2010, fiore2017}. These outflows are believed to deposit large amounts of energy and momentum into the interstellar medium \citep[e.g.][]{zubovas2012} and affect the host galaxy gas reservoir available for both star formation and SMBH accretion, offering a possible explanation for the observed $M_\mathrm{BH}$$-$$\sigma$ relation \citep[e.g.][]{ferrarese2000}. Accordingly, the AGN feedback mechanism should manifest itself in full force in high-\ledd\ sources, making them the ideal laboratory for probing the real impact of nuclear activity on the evolution of massive galaxies \citep[e.g.][]{reeves2009, nardini2015, tombesi2015, marziani2016, bischetti2017, bischetti2019, laurenti2021}.

\section{Correlation with X-ray properties}\label{x-ray}

X-ray observations provide the invaluable opportunity to investigate the emission arising from the nuclear region of AGN, and have often concurred to highlight interesting correlations between the supermassive black hole accretion properties and various X-ray parameters.
For instance, a long-standing issue consists in the existence of a positive correlation between the photon index of the X-ray power-law continuum $\Gamma$ and \ledd, implying a very steep continuum (i.e. $\Gamma>2$) is a prerogative of AGN with \ledd$\,\gtrsim0.3$ \citep[e.g.][]{shemmer2008, risaliti2009, brightman2013, fanali2013, huang2020, liu2021}.
This is often explained in terms of an enhanced UV emission from the disc, that improves the efficiency of the radiative cooling in the X-ray corona and, consequently, leads to the reduction of the electron temperature, resulting in an overall steepening of the power-law continuum.
The discovery of such a strong dependence of $\Gamma$ on $\log$\ledd (i.e. $\Gamma \sim 0.3\,\times\,\log\lambda_\mathrm{Edd} + 2$; e.g. \citet{shemmer2008, risaliti2009, brightman2013}) elicited immediate interest due to its possible application in X-ray extragalactic surveys, as it would allow to estimate the $M_\mathrm{BH}$ for AGN with a reliable measurement of $\Gamma$, including type-2 AGN, for which commonly used ‘single-epoch’ $M_\mathrm{BH}$ estimators are not applicable.
However, \citet{trakhtenbrot2017} questioned the existence of a strong correlation between $\Gamma$ and \ledd\, reporting only a very weak correlation after analysing a large sample of low-$z$ ($0.01 \!<\! z \!<\! 0.5$) sources with broadband X-ray spectra from the BAT AGN Spectroscopic Survey.

Furthermore, \citet{lusso2010} studied the hard X-ray bolometric correction, $k_\mathrm{bol,X}\!=\! L_\mathrm{bol}/L_\mathrm{2-10\,keV}$, and the optical/UV-to-X-ray spectral slope ($\alpha_\mathrm{ox}$) of AGN in the COSMOS survey as a function of $\log$\ledd, and reported on a positive correlation between the Eddington ratio and the two X-ray quantities.
These relations support a scenario in which the accretion disc-corona system of highly accreting AGN is different from that of standard AGN, producing a relatively weaker X-ray emission as compared to the optical/UV.
However, a full understanding of the correlations involving \ledd, $\Gamma$ and the other X-ray parameters is severely hampered by the lack of a large number of high-\ledd\ AGN, raising the challenging task of increasing the amount of data related to AGN belonging to the extreme high-end tail of the \ledd\ distribution. 

\section{X-HESS}\label{x-hess}

As discussed in Sect.\ \ref{intro}, except for some local NLSy1, there are only a few pointed observations of high-\ledd\ AGN especially at higher $z$.  
Thus if we merely rely on such dedicated observations, then building a sample of highly accreting AGN which is both statistically sound and well representative of the whole population is extremely difficult. 
However, many of them can be serendipitously found in the field of other X-ray sources which have been intensively observed thanks to deep observational campaigns carried out with \textit{XMM-Newton} \citep[][]{jansen2001}.
For this reason, the analysis of serendipitous high-\ledd\ AGN discloses the unprecedented possibility to investigate not only their spectral but also variability properties in the X-rays in a much wider range of black hole mass, bolometric luminosity and redshift.

In order to define a large sample of high-\ledd\ serendipitous AGN, we choose to exploit the extended database included in the 10th release of the \textit{XMM-Newton} Serendipitous Source Catalogue (4XMM-DR10; \citet{webb2020}), that contains $\sim11600$ observations in which approximately 850000 sources have been detected. 
To recover only those X-ray observations of spectroscopically-confirmed AGN, we consider the catalogue of spectral properties of QSOs from the Sloan Digital Sky Survey Data Release 14 \citep[][]{rakshit2020}, including $\sim526000$ AGN, and crossmatch it with the serendipitous \textit{XMM-Newton} catalogue within a radius of 3 arcsec in coordinates  to avoid possible spurious identifications, while maximising the completeness of the sample.
Then we impose the additional requirements that (i) each source with multi-epoch detections must have at least one good quality observation in terms of photon counts in the broad ($E\!=\!0.2\!-\!12$ keV) EP8 band, i.e. EP\_8\_CTS $\!\geq \! 1000$, to ensure robust spectral results and (ii) consider only those AGN with a sufficiently high value of $\log\,$\ledd\ that we set to $\log\,$\ledd\ $\!\geq\!-0.2$.
At this stage, we obtain a sample of 88 AGN for a total of 202 observations. Since we are only interested in radio-quiet AGN, we remove from the sample seven sources that have been classified as radio-loud and also three more quasars that are known to be lensed. 
Starting from this sample of 78 AGN, we impose the further condition that the $M_\mathrm{BH}$ estimate of each source must be derived from the full width at half maximum (FWHM) of either H{\scriptsize{$\beta$}} or Mg{\,\scriptsize{II}} emission lines, which are more reliable estimators than the FWHM of C{\,\scriptsize{IV}} \citep[e.g.][]{baskin2005, coatman2017, vietri2018, vietri2020}.

\begin{figure}[t]
\centerline{\includegraphics[width=.84\columnwidth]{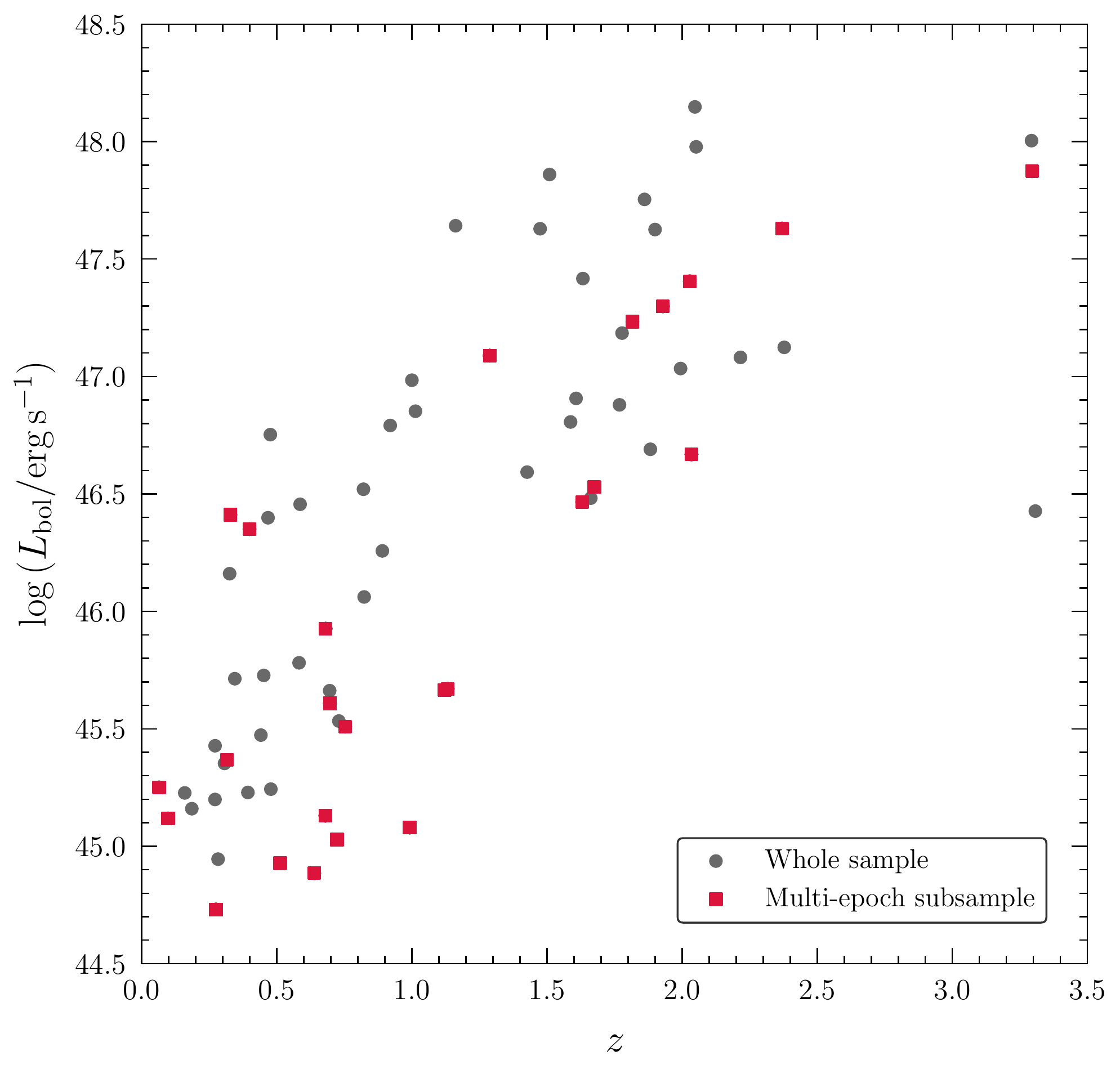}}
\caption{\scriptsize{Distribution of the X-HESS sources in the $\log{L_\mathrm{bol}}-z$  plane. AGN with individual or multi-epoch observations are shown in grey and red, respectively. The multi-epoch subsample is well representative of the whole sample in terms of these parameters.}}
\label{fig:L_z}
\end{figure}

Finally, we obtain a catalogue of 69 AGN with 161 observations that we name as \textit{XMM-Newton} High-Eddington Serendipitous AGN Sample, or X-HESS.
\begin{figure}[t]
\centerline{\includegraphics[width=.8\columnwidth]{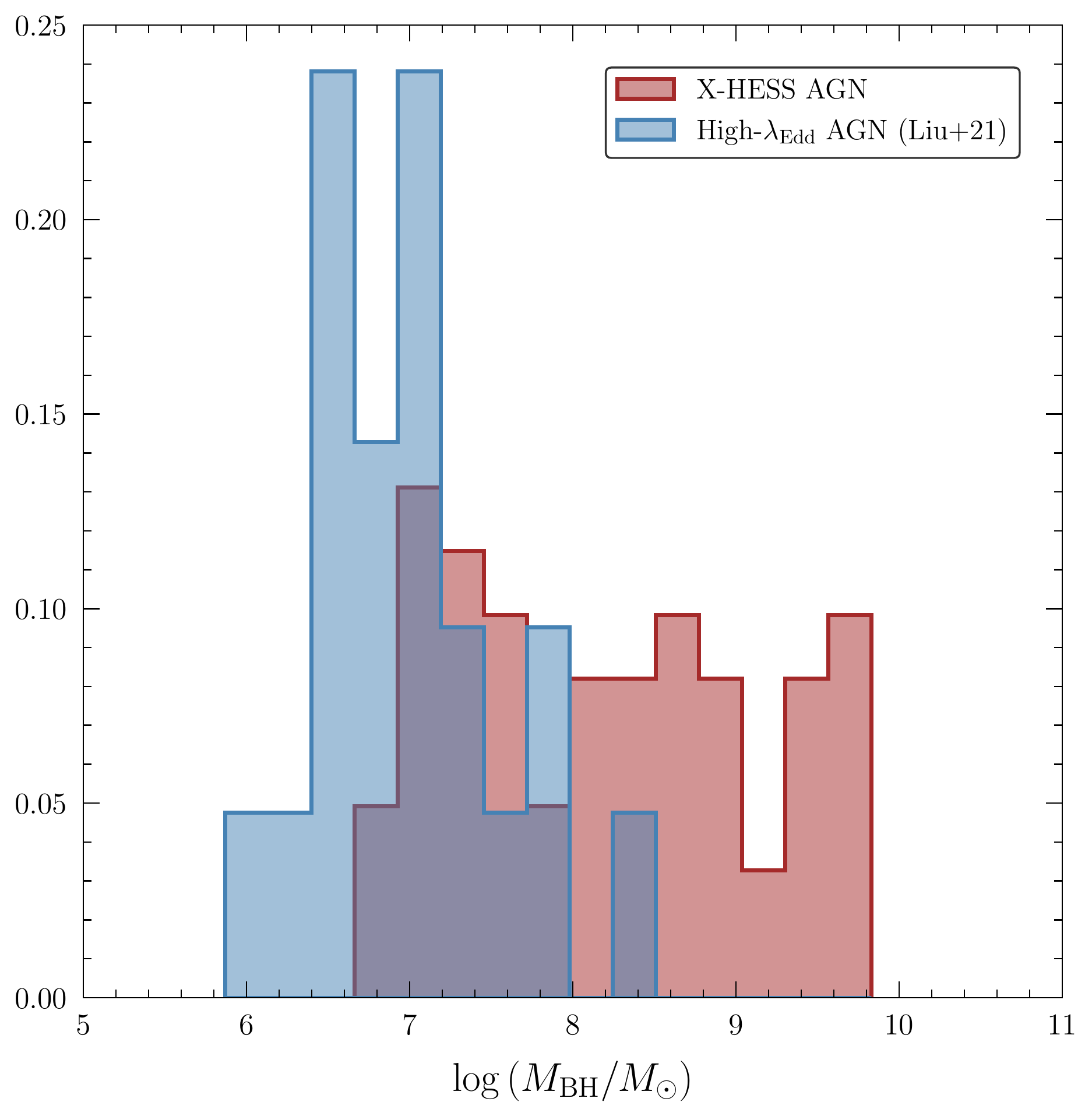}}
\caption{\scriptsize{Relative frequency histogram of $M_\mathrm{BH}$ in the X-HESS sample (in red) and the subsample of highly accreting AGN (in blue) analysed by \citet{liu2021}. X-HESS allows us to extend the study of high-\ledd\ AGN towards the poorly-explored extreme end of the black hole mass distribution.}}
\label{fig:histMbh}
\end{figure}
The X-HESS AGN are distributed over wide intervals of redshift (0\,<$\,z\,$<\,3.3), bolometric luminosity (44.7\,<$\, \log(L_\mathrm{bol}/\mathrm{erg\,s}^{-1})\,$<\,48.1), black hole mass (6.8\,<$\, \log({M_\mathrm{BH}/M_\odot})\,$<\,9.9) and Eddington ratio ( $-0.2\!<\! \log\lambda_\mathrm{Edd}\!<\! 0.78$ ). 
The presence of AGN spanning a wide range of $M_\mathrm{BH}$, $L_\mathrm{bol}$ and $z$ contributes to make X-HESS a useful tool to improve our understanding of high-\ledd\ AGN in those regions of the parameter space of the main physical quantities that have been poorly sampled so far.
Specifically, the broad $M_\mathrm{BH}$ distribution of the X-HESS AGN (see Fig.\ \ref{fig:histMbh}) allows us to extend the analysis of highly accreting sources towards AGN with black hole mass values substantially higher than those previously considered in other studies of high-\ledd\ quasars in the local Universe \citep[e.g.][]{lu2021,liu2021}.
Reported in Fig.\ \ref{fig:L_z} is the $\log{L_\mathrm{bol}}-z$ distribution of the X-HESS sources, where AGN with individual or multi-epoch observations are described by points highlighted in different colours. The multi-epoch subsample of X-HESS is well representative of the whole sample in terms of the main physical parameters, including $M_\mathrm{BH}$ and \ledd\, and offers the unprecedented opportunity to investigate the X-ray flux and spectral variability of 25 highly accreting sources which have been repeatedly observed by \textit{XMM-Newton} for 117 times.
For this reason, as a first step we are currently concentrating our efforts to analyse this subsample.
Furthemore, this study can be complemented with optical/UV data, allowing us to investigate the interplay with the X-rays (e.g. with $\alpha_\mathrm{ox}$), by taking advantage of the observations carried out by the Optical Monitor \citep[OM; ][]{mason2001} aboard \textit{XMM-Newton}. Indeed, the vast majority of the X-HESS sources ($\sim70\%$) can rely on simultaneous OM observations obtained by crossmatching X-HESS with the latest release of the XMM-OM Serendipitous Ultraviolet Source Survey Catalogue \citep[SUSS5.0; ][]{page2012}.
This could allow us e.g.\ to study correlations between X-ray and optical/UV on timescales longer than the light-travel time between the disc and corona, both within single observations (for the longest exposures $\sim\!100$ ks) as well as between different observations, typically separated by times from weeks to years.
The heterogeneous properties of the X-HESS sources, including the multi-epoch AGN, allow us to extend the dynamic range of the relations involving \ledd\ and X-ray spectral parameters, such as the one involving the photon index $\Gamma$ discussed in Sect.\ \ref{intro}, by significantly populating the poorly explored extreme end of the \ledd\ distribution with AGN covering a large parameter space in terms of $z$, $M_\mathrm{BH}$ and $L_\mathrm{bol}$.

\begin{figure}[t]
\centerline{\includegraphics[width=.84\columnwidth]{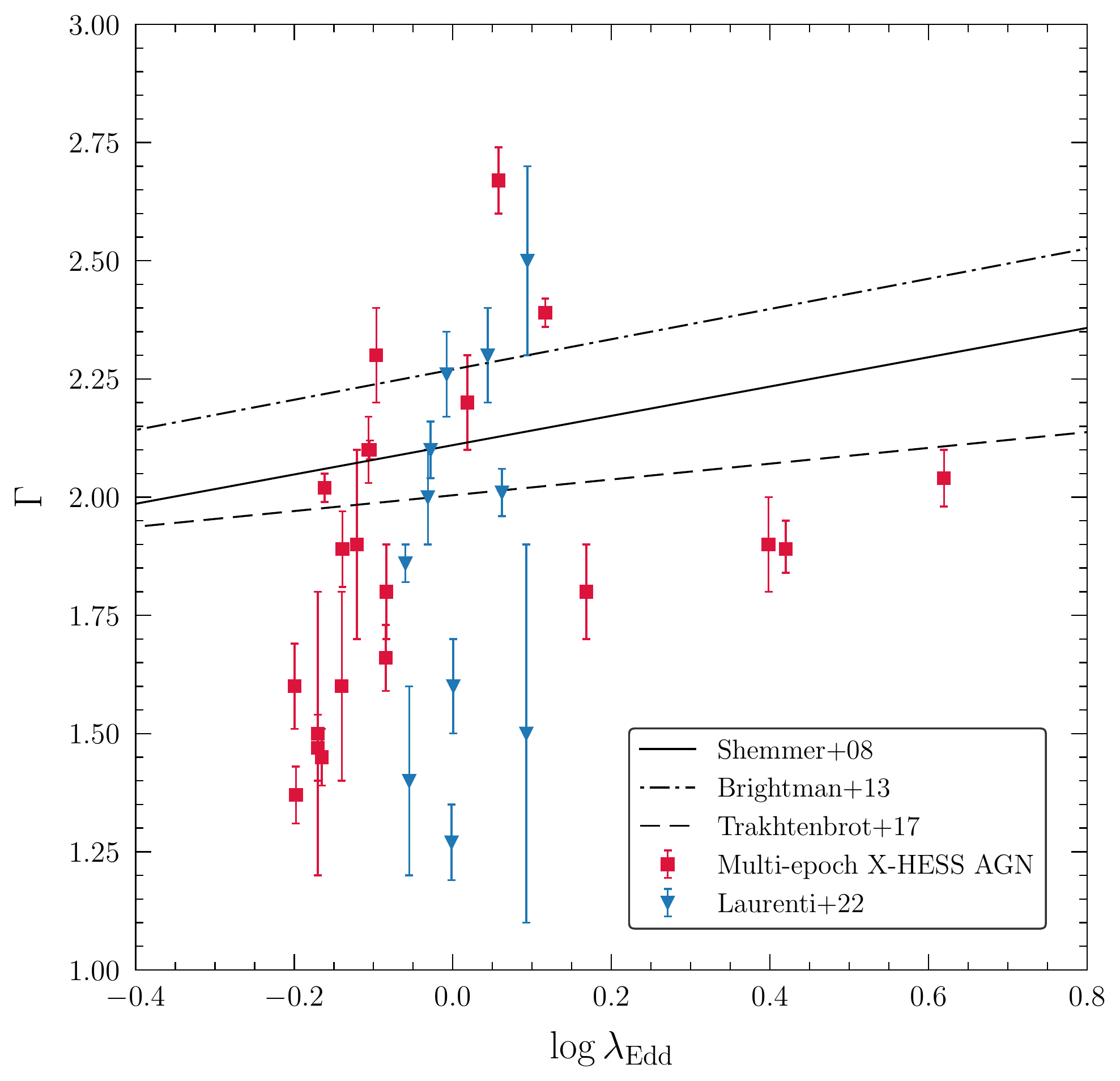}}
\caption{\scriptsize{$\Gamma$ vs $\log\,$\ledd\ for the highly accreting AGN in the multi-epoch X-HESS subsample (in red). High-\ledd\ AGN from our previous work \citet{laurenti2022} are also shown in the same figure (blue triangles). Solid, dashed and dash-dotted lines describe the expected relation from other studies. }}
\label{fig:Gam_Edd}
\end{figure}

Shown in Fig.\ \ref{fig:Gam_Edd} is the distribution of the multi-epoch X-HESS AGN in the $\Gamma$ vs $\log\,$\ledd\ plane, where the photon index is derived for each source from a complete one-by-one spectral analysis, adopting the same procedure described in \cite{laurenti2022}, associated to the observation with the longest exposure. 
The expected relations from other studies are plotted in the same figure, where we also include the points describing the high-\ledd\ AGN sample analysed in our previous work \citet{laurenti2022}. 
Regarding the X-HESS AGN, given the energy dependence of the effective area of \textit{XMM-Newton}, the selection criterion based on the number of counts in the EP8 band is approximately translated into a condition about the photon counts in the soft X-rays, as the counts above $>\!2$ keV are expected to give a minor contribution. Thus, in principle, this might introduce a bias due to the fact that we are likely to select preferentially sources with a soft X-ray continuum, i.e.\ with large values of $\Gamma$.
However, the preliminary results from X-HESS are indicative of a large dispersion of $\Gamma$ values in the $\Gamma-\log\,$\ledd\ plane, with a conspicuous number of sources showing photon indices less steep than expected. This evidence is apparently in agreement with our previous findings, thus supporting the difficulty to utilise the expected relations to obtain an accurate measurement of the black hole mass from the X-ray spectrum.
This may also indicate that the interplay of the accretion disc-corona system in such highly accreting sources is more complicated than what previously thought and might affect the slope of the X-ray emission by causing a redistribution (perhaps time-dependent) of the disc accretion power dissipated in the corona, such as a variation in the inner truncation radius of the accretion disc or the geometry of the disc-corona complex \citep[e.g.][]{kubota2018}.

\begin{figure}[t]
\centerline{\includegraphics[width=\columnwidth]{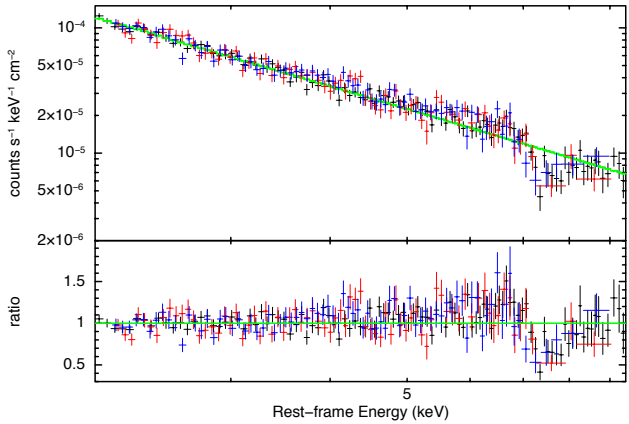}}
\caption{\scriptsize{Hard ($E>2$ keV rest frame) X-ray spectrum of PG 1448+273 modelled with a power law modified by Galactic absorption (in green). Residuals with respect to this simple model are shown in the bottom panel. Data from EPIC-pn, MOS1 and MOS2 are shown in black, red and blue, respectively. The UFO absorption feature at $E\sim7.5$ keV is clearly visible in both panels. }}
\label{fig:PG}
\end{figure}

Despite the preliminary status of the analysis of the multi-epoch X-HESS AGN, some individual sources have already provided interesting case studies. For instance, in one of the two \textit{XMM-Newton} observations of the NLSy1 galaxy PG 1448+273 ($z=0.0645$) we found evidence of an absorption feature at $E\sim7.5$ keV (see Fig.\ \ref{fig:PG}) which represents the spectral imprint of a powerful UFO, that we analysed in \citet{laurenti2021} by means of a novel photoionisation model called Wind in the Ionised Nuclear Environment \citep[WINE; ][]{luminari2018}.

\begin{figure}[ht]
\centerline{\includegraphics[width=.84\columnwidth]{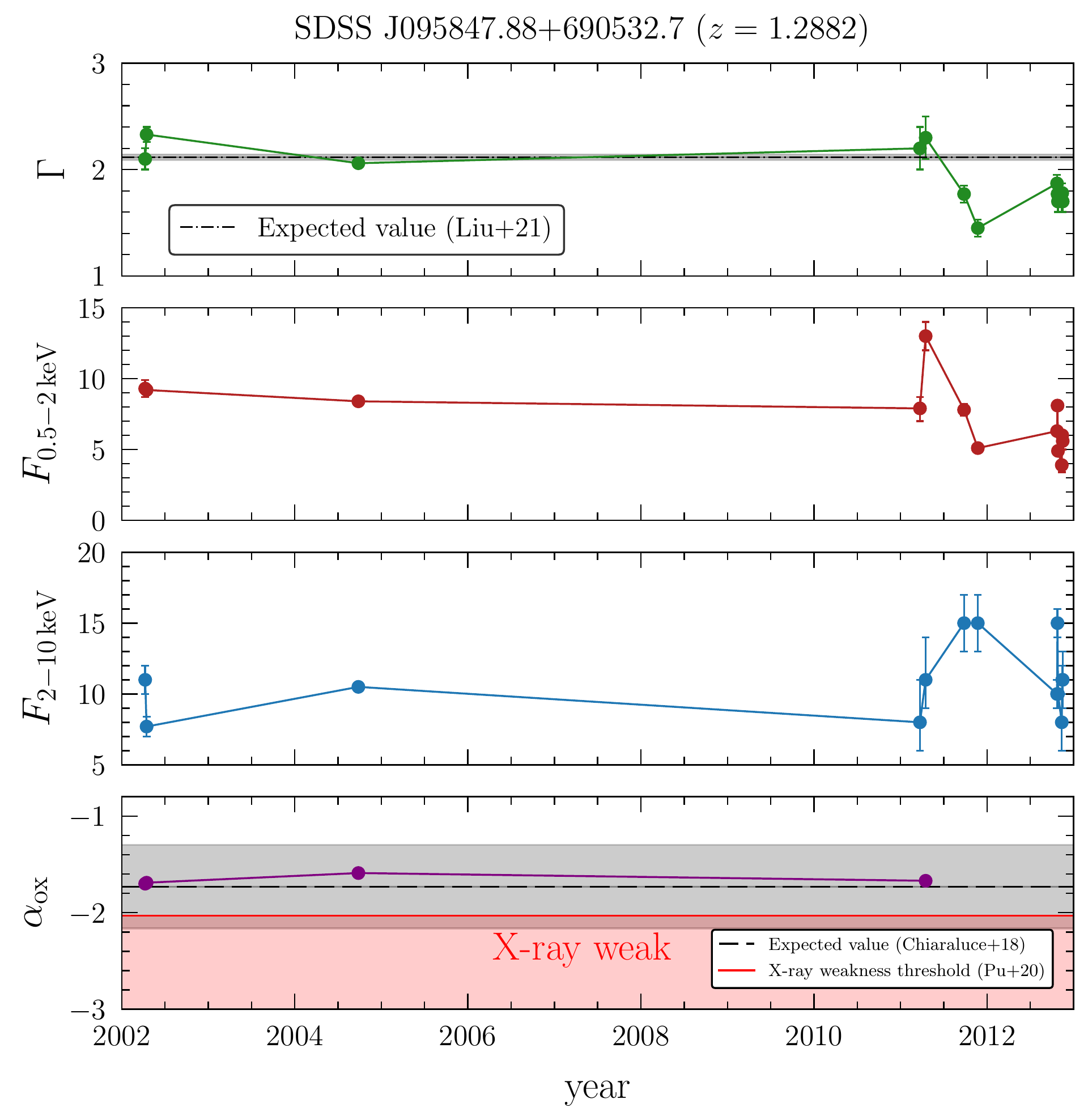}}
\caption{\scriptsize{Light curves of the photon index $\Gamma$, soft X-ray flux $F_\mathrm{0.5-2\,keV}$, hard X-ray flux $F_\mathrm{2-10\,keV}$ and $\alpha_\mathrm{ox}$ for the X-HESS AGN SDSS J095847.88+690532.7. Fluxes are considered in units of $10^{-14}$ erg cm$^{-2}$ s$^{-1}$. Remarkable flux variations are seen over timescales as short as approximately one week in the rest frame of the source. The expected values of $\Gamma$ and $\alpha_\mathrm{ox}$ from the $\Gamma-\log{\lambda_\mathrm{Edd}}$ and $\alpha_\mathrm{ox}-\log{L_\mathrm{UV}}$ relations of \citet{liu2021} and \citet{chiaraluce2018} are also shown in the top and bottom panels, respectively. The threshold for X-ray weakness as defined by \citet{pu2020} is also displayed in the bottom panel.}}
\label{fig:2_LC}
\end{figure}

In addition, X-HESS does also include sources that show remarkable X-ray flux variations over different timescales. As an example, SDSS J095847.88+690532.7, one of the most densely observed X-HESS AGN hosting a $\gtrsim\,$$10^9\,M_\odot$ black hole at redshift $z\!\sim\! 1.3$ shows large variations in terms of $\Gamma$ and both soft ($E=0.5\!-\!2$ keV) and hard ($E=2\!-\!10$ keV) X-ray flux over more than a decade, as displayed in Fig.\ \ref{fig:2_LC}.
We find flux variations of a factor of $> 2$ in approximately one week in the rest frame of the source, which are relatively high for an AGN with such a massive SMBH.
The values of $\Gamma$ in Fig.\ \ref{fig:2_LC} range between $\sim\!1.5$ and $\sim\!2.3$, indicating a large dispersion around the expected value from the adopted $\Gamma-\log\lambda_\mathrm{Edd}$ relation, in this case the one from \citet{liu2021}, can be seen also in individual sources when repeatedly observed.

\section{Conclusion}

High-\ledd\ AGN (i.e.\ \ledd\ $\gtrsim$ 0.5) represent interesting case studies because of their peculiar nuclear accretion properties, wind-acceleration mechanisms and cosmological implications, but due to their relative paucity in the local Universe they have often been overlooked. Moreover, firm conclusions on hotly-debated correlations involving \ledd\ and many X-ray parameters are severely hampered by the lack of a substantial number of observations of AGN in the high-end tail of the \ledd\ distribution. 
To overcome this limitation we take advantage of the data included in the 10th release of the \textit{XMM-Newton} Serendipitous Source Catalog, which allow us to build a sample of 69 high-\ledd\ AGN with serendipitous X-ray observations, covering large intervals of $z$, $M_\mathrm{BH}$ and $L_\mathrm{bol}$ that we called X-HESS. The X-ray data are also complemented with the simultaneous optical/UV observations from the latest release of the serendipitous OM catalogue. Preliminary results from X-HESS indicate that:

\begin{itemize}
\item The $\Gamma-\log\lambda_\mathrm{Edd}$ distribution of high-\ledd\ AGN is characterised by a large scatter in terms of $\Gamma$, showing values of the photon index less steep than those expected from previously reported correlations \citep[e.g.][]{shemmer2008, risaliti2009, brightman2013, liu2021}, in agreement with what we found in our recent X-ray spectroscopic analysis of a sample of fourteen high-\ledd\ AGN \citet{laurenti2022}. 

\item Remarkable spectral and flux variations are observed in the first X-HESS sources we have already analysed. In the case of PG 1148+273 we find the spectral imprint of a powerful UFO \citet{laurenti2021}, while for SDSS J095847.88+690532.7, one of the most densely monitored X-HESS AGN, we see considerable soft and hard X-ray flux variations over timescales as short as almost one week in the rest frame of the source.

\end{itemize}

\noindent In the near future we plan to complete the analysis of the whole X-HESS sample, starting from the multi-epoch subsample, which would allow us to improve our understanding of highly accreting AGN by investigating both the spectral and flux variations of 25 high-\ledd\ sources over different timescales, from a few days to several years. 
%As the average black hole mass of the X-HESS AGN is about $\sim\!10^8\,M_\odot$ and assuming a distance of $10\,R_\mathrm{S}$, with $R_\mathrm{S}\!=\! 2GM_\mathrm{BH}/c^2$ being the Schwarzschild radius, between the X-ray corona and the inner accretion disc, we find a characteristic light-crossing time of $t_\mathrm{lc}\!\sim\!10$ ks. Since $t_\mathrm{lc}$ is shorter, on average, than the net exposure of the \textit{XMM-Newton} observations, we could probe the possible correlation between the X-ray and optical/UV variability and study the interplay of the accretion disc-corona complex by using both EPIC and MOS data, when available.  
We also plan to upgrade X-HESS by considering the current latest release of the \textit{XMM-Newton} serendipitous catalogue (4XMM-DR12) and take advantage of each future release as well to keep our sample always up-to-date.          

\backmatter

\section*{Acknowledgments}

We thank the anonymous referee for her/his useful comments.

%\nocite{*} % Show all bib entries - both cited and uncited; comment this line to view only cited bib entries;
\bibliography{X-HESS}%

\end{document}